%% file: main.tex
\documentclass[sigconf]{acmart}

\usepackage{enumitem}
\usepackage{tikz}
\usepackage{array}

\usetikzlibrary{external, shapes.geometric, arrows}
\usepackage{xspace}
\acmDOI{}          
\acmISBN{}
 \title[short]{full}

\AtBeginDocument{%
  \providecommand\BibTeX{{%
    \normalfont B\kern-0.5em{\scshape i\kern-0.25em b}\kern-0.8em\TeX}}}

\copyrightyear{2026}
\acmYear{2026}
\setcopyright{cc}
\setcctype{by}
\acmConference[UIST Adjunct '26]{The 39th Annual ACM Symposium on User Interface Software and Technology}{November 02--05, 2026}{Detroit, MI, USA}
\acmBooktitle{The 39th Annual ACM Symposium on User Interface Software and Technology (UIST Adjunct '26), November 02--05, 2026, Detroit, MI, USA}
\acmDOI{10.1145/3830397.3841822}
\acmISBN{979-8-4007-2855-6/2026/11}

\begin{document}

\tikzstyle{5_box_node} = [
    rectangle,
    rounded corners, 
    minimum width=2cm, 
    minimum height=1cm,
    text centered,
    text width = 2.5cm,
    draw=black,
]
\tikzstyle{3_box_node} = [
    rectangle,
    rounded corners, 
    minimum width=3cm, 
    minimum height=1cm,
    text centered,
    text width = 4cm,
    draw=black,
]
\tikzstyle{4_box_node} = [
    rectangle,
    rounded corners, 
    minimum width=3cm, 
    minimum height=1cm,
    text centered,
    text width = 3.2cm,
    draw=black,
]
\tikzstyle{arrow} = [thick,->,>=stealth]


\newcommand{\yaqing}[1]{\textcolor{violet}{#1}}
\newcommand{\sssec}[1]{\vspace*{0.05in}\noindent\textbf{#1}}

\newcommand{\sys}{\text{ReVision}\xspace}

\title{ReVision: Supporting Designers' Interpretation and Exploration of Visuals in Concepts and Forms}

\author{Yaqing Yang}
\authornote{Both authors contributed equally to this research.}
\affiliation{%
  \institution{Carnegie Mellon University}
  \city{Pittsburgh, PA}
  \country{USA}}
\email{yaqingyy@andrew.cmu.edu}

\author{Mei-Xi Chia}
\authornotemark[1]
\affiliation{%
  \institution{Carnegie Mellon University}
  \city{Pittsburgh, PA}
  \country{USA}}
\email{mchia@andrew.cmu.edu}

\author{Aniket Kittur}
\affiliation{%
  \institution{Carnegie Mellon University}
  \city{Pittsburgh, PA}
  \country{USA}}
\email{nkittur@cs.cmu.edu}

\renewcommand{\shortauthors}{Yang et al.}
\begin{teaserfigure}
  \includegraphics[width=\textwidth]{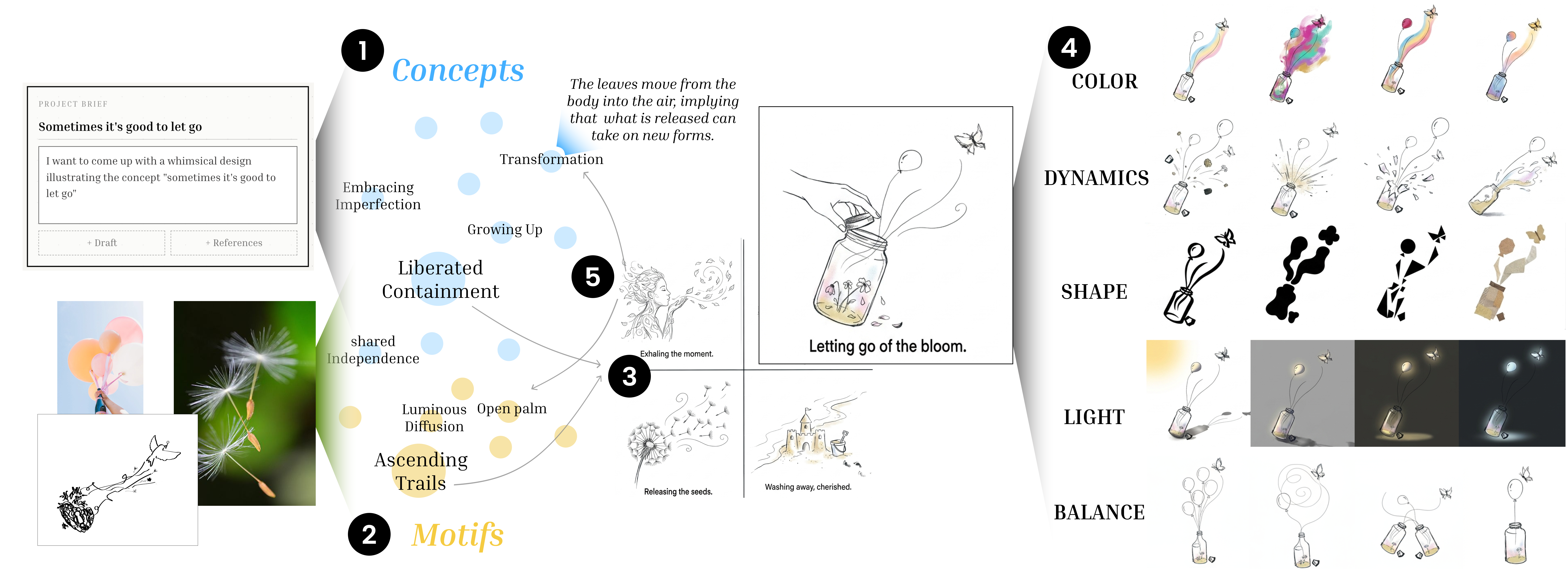}
  \caption{ReVision supports visual designers' broad exploration of ideas by decomposing briefs and references\protect\footnotemark[1] into (1) concepts and (2) motifs, (3) recombining them into new visual directions, (4) systematically varying their visual forms, and (5) feeding generated visuals back into the exploration space.}
  \label{fig:teaser}
\end{teaserfigure}

\include{parts/p0-abstract}


\maketitle

\input{parts/p1-intro}

\footnotetext[1]{Reference images used in Figure 1 are from public available resources: 
\url{https://www.skyboxeventproductions.com/veuve-clicquot-rose/} and \url{https://www.flickr.com/groups/macroquality/pool/}}

\input{parts/p3-system}
\input{parts/p4-futurework}

\bibliographystyle{ACM-Reference-Format}
\bibliography{sample-base}

\end{document}

%% file: parts/p0-abstract.tex
\begin{abstract}
Visual designers get inspiration from references to expand their design space. They decompose what makes a reference evocative into conceptual and visual elements, ranging from explicit attributes such as objects and colors, to less readily articulated concepts and visual motifs. They then create different visual forms to explore how the selected elements could be combined differently. Novices often struggle with these moves, instead focusing on surface features or producing limited visual variation, thus becoming fixated on the reference. Existing tools support editable visual attributes and high-level themes, but provide limited control over how conceptual interpretations relate to expressive visual motifs or how their combinations can be systematically re-expressed. We present ReVision, an AI-based tool that decomposes visual and textual references into editable conceptual interpretations and visual motifs, enables their recombination across conceptual and visual spaces, and renders each direction as divergent visual-form variations, supporting more divergent exploration during the creation process.
\end{abstract}

\keywords{Human-AI Collaboration, Generative AI, Creativity Support}

%% file: parts/p1-intro.tex
\section{Introduction}
\label{intro} 
Visual references can expand a designer's thinking, but they can also constrain it when designers focus only on surface features such as what a reference depicts~\cite{jansson1991design}. A key design skill is to reinterpret references by identifying what makes them evocative across both conceptual and visual space: what the reference conveys and how it conveys it. Conceptual elements may range from literal descriptions of what appears in an image to higher-level interpretations (e.g., liberation, transformation). Visual motifs may range from identifiable objects, colors, and styles to abstract structures (e.g., ascending trails, diffused light). 
Importantly, the same conceptual interpretation can be expressed through different visual motifs, and the same visual motif can support different conceptual interpretations depending on how it is used.
Recognizing and manipulating this relationship helps designers move beyond surface copying, retain a conceptual interpretation while changing its visual motif, or reuse a visual motif toward a different conceptual interpretation~\cite{oxman2002thinking}. Designers can then detach and recombine these elements into new visual ideas, and explore the ideas through variations in visual form, such as composition, color, light, or movement~\cite{arnheim1954art}. Novices often struggle with those moves,
instead reproducing surface features and committing prematurely to limited visual expressions~\cite{oxman2002thinking,yen2024processgallery}.

Prior systems represent references through editable visual attributes~\cite{shi2025brickify,choi2024creativeconnect}, themes~\cite{choi2026ideablocks,riche2025ai}, latent visual features without text~\cite{vinker2023concept}, or blends based on object features or related contexts~\cite{chilton2019visiblends,wang2023popblends}. However, these approaches provide limited support for making explicit how a reference can be interpreted through both conceptual interpretations and visual motifs, and for turning these interpretations into editable materials for further exploration. This can make it difficult for designers to move beyond surface features and reuse what makes a reference evocative in new directions. Without decomposition, these interpretations remain implicit; without recombination, they remain descriptive; and without systematic variation, a new direction may become fixed in its first visual expression.

To fill this gap, we present \sys, a system that turns references into composable material for visual ideation. \sys decomposes visual references or textual themes into editable conceptual interpretations and visual motifs, enables designers to revise and recombine them across conceptual and visual spaces, and generates variants of each resulting image across 5 dimensions of visual forms (e.g., balance, shape). These visuals can be interpreted and recombined again, supporting iterative and divergent exploration.

%% file: parts/p3-system.tex
\section{System Walkthrough}

\sys supports broad visual exploration by turning references into reusable concepts and motifs, and generating diverse variations from selected combinations of these elements. The system’s main features are mainly powered by a large language model, a vision-language model, and a text-to-image generation model.

\begin{figure}[h]
\vspace{-8pt}
  \centering
\includegraphics[width=\linewidth]{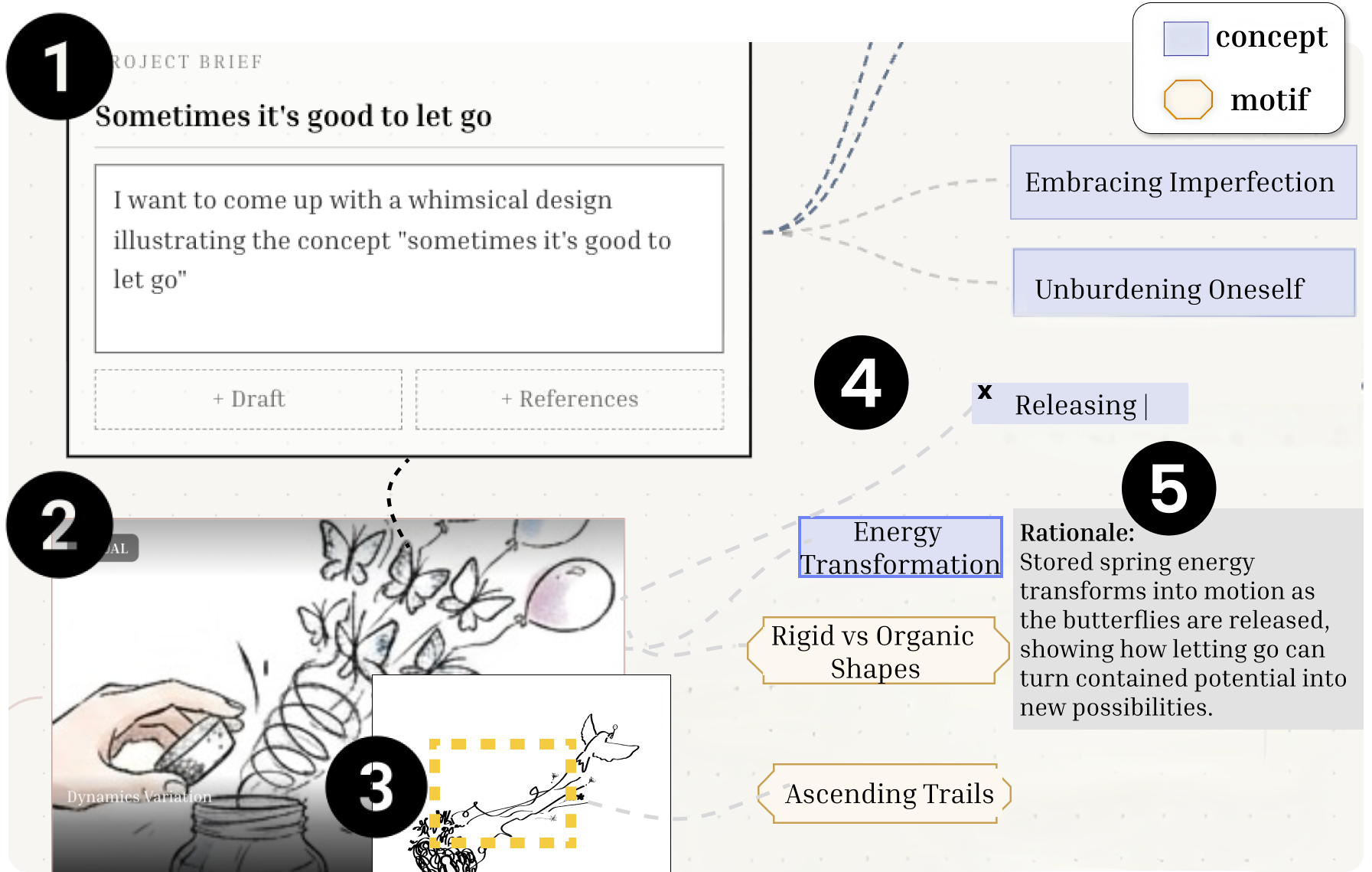}
  \caption{\sys interface for decomposing (1) design briefs and (2) visual references into concepts and motifs, with (5) explanations of their connection to the reference. Users can also (3) select image regions to analyze and (4) add or edit extracted concepts and motifs.}
  \Description{interface screenshot shows how \sys decomposes (1) design briefs and (2) visual references into concepts and motifs. Users can also (3) select a specific part of an image to analyze and (4) add or edit concepts and motifs.
}
\vspace{-8pt}
  \label{fig:decompose}
\end{figure}

\sssec{Decomposition: Turning Briefs and References into Interpretive Materials.}
To help designers identify what makes a reference evocative, \sys decomposes it into two types of editable materials. \textit{Conceptual interpretations} describe what the reference expresses, ranging from literal descriptions to less readily articulated concepts. For example, in Figure~\ref{fig:decompose}, `releasing' represents a specific, literal action, whereas `embracing imperfection' represents a more abstract concept. \textit{Visual motifs} describe how these meanings are conveyed, ranging from common visual dimensions such as objects, colors, and styles to expressive structures that are difficult to capture through fixed dimensions. One example is the \textit{rigid organic shape} in Figure~\ref{fig:decompose}, which expresses the tension between the rigid form of a glass bottle and the curved forms created by the butterflies and their movement trajectories. Separating what a reference expresses from how it expresses it helps designers move beyond copying surface features. As shown in Figure~\ref{fig:decompose}(5), \sys explains how an extracted concept is grounded in specific visual evidence, making the connection between conceptual interpretation and visual expression visible to users. Designers can then revise, retain, or reuse either type of material in a different direction. As shown in Figure~\ref{fig:decompose}, designers can also edit or add interpretations and select a specific image region (Figure~\ref{fig:decompose}, (3)(4)) for further analysis.

\sssec{Combination: Constructing New Designs across Conceptual and Visual Space.}
For combination process, designers can select concepts, motifs, and visuals flexibly to generate new visuals in \sys (Figure~\ref{fig:teaser}(3)). As exploration proceeds, generated concepts, motifs, and visuals remain as traces that designers can cluster under self-defined themes. These organized traces make exploration histories available for continued design exploration~\cite{marquardt2025imaginationvellum,ling2026inspirationgraph}, while helping designers identify relationships among concepts and visuals and build new combinations on top of the evolving trace graph.


\begin{figure}[h]
  \vspace{-8pt}
  \centering
 \includegraphics[width=\linewidth]{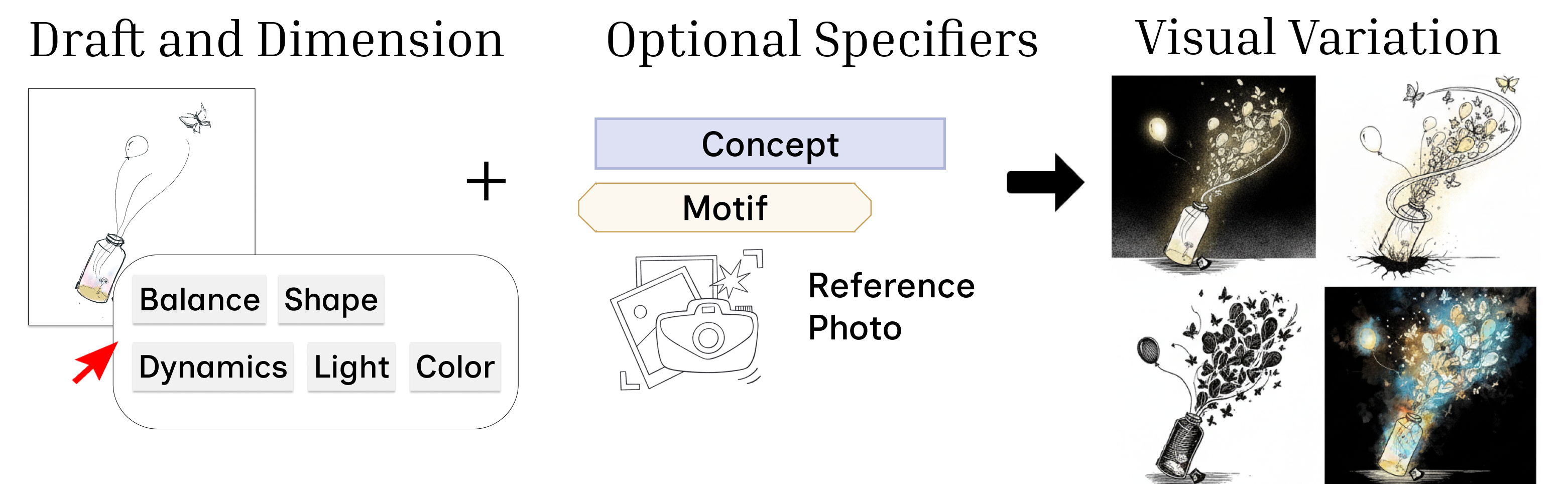}
  \caption{\sys combines concepts, motifs, and an optional reference image to create visual variations of a draft along specified dimensions.}
  \label{fig:combine2}
  \Description{\sys combines concepts, motifs, and an optional reference image to create visual variations of a draft along specified dimensions.}
  \vspace{-8pt}
\end{figure}

\sssec{Variation: Systematically Re-expressing an idea.}
A generated image through combination feature represents only one visual expression of the selected concepts and motifs, presenting a single image as the result may lead designers to fixate on the first result. Simply rerunning the same prompt does not help: in our initial tests, it produced either homogeneous or unrelated images because it did not vary any specific aspect. To support more deliberate exploration, \sys generates systematic variants from a seed image (Figure~\ref{fig:combine2}). Drawing on Arnheim's theory of visual perception~\cite{arnheim1954art}, we define 5 visual forms: 1) \textit{balance}: how visual weight is distributed to create equilibrium or instability; 2) \textit{shape}: how geometric and organic forms structure what is perceived; 3) \textit{light}: how illumination, shadow, and luminance create depth and volume; 4) \textit{color}: how hues, saturation, and contrast shape mood and symbolic meaning; and 5) \textit{dynamics}: how movement, tension, and visual energy are expressed. \sys conditions the image generation model on the seed image, the related concepts, and the chosen visual form. It then generates variants that changes that form while preserving the underlying design direction (Figure~\ref{fig:teaser}(4)). This supports designers to compare different visual expressions of the same conceptual-visual materials.

%% file: parts/p4-futurework.tex
\section{Conclusion and Future Work}
We present \sys, an AI-based tool that supports designers in exploring design directions across the conceptual and visual spaces. By decomposing references into manipulable elements, recombining them into new visuals, and systematically varying their visual form, \sys makes interpretation concrete and supports more divergent exploration of the design space. In future work, we plan to formally evaluate \sys in user studies to understand whether externalizing concepts and motifs shifts novices from surface replication toward conceptual transformation.